\documentclass[sigconf]{acmart}

\usepackage{adjustbox}
\usepackage{bm}
\usepackage{booktabs}
\usepackage{caption}
\usepackage{color}
\usepackage{makecell}
\usepackage{microtype}
\usepackage{multirow}
\usepackage{paralist}
\usepackage{subcaption}
\usepackage{tabularx}
\usepackage{url}
\usepackage{xcolor}

\PassOptionsToPackage{hyphens}{url}\usepackage{hyperref} 
\definecolor{mylinkcolor}{RGB}{0,0,0}
\hypersetup{colorlinks,allcolors=mylinkcolor,citecolor=mylinkcolor}
\usepackage[capitalize]{cleveref}

\newcommand{\arbnote}[1]{\textcolor{magenta}{ARB: #1}}
\renewcommand{\arbnote}[1]{}

\newcommand{\omt}[1]{}

\title{Frozen Binomials on the Web: \\ Word Ordering and Language Conventions in Online Text}

\newcommand{\xhdr}[1]{\vspace{0.5mm}\noindent{\textbf{#1.}}\hspace{0.5mm}}

\author{Katherine Van Koevering}
\email{kav64@cornell.edu}
\affiliation{
  \institution{Cornell University}
}

\author{Austin R.\ Benson}
\email{arb@cs.cornell.edu}
\affiliation{
  \institution{Cornell University}
}

\author{Jon Kleinberg}
\email{kleinberg@cornell.edu}
\affiliation{
  \institution{Cornell University}
}

\begin{document}

\begin{abstract}
There is inherent information captured in the order in which we write words in a list. The orderings of binomials --- lists of two words separated by `and' or `or' --- has been studied for more than a century. These binomials are common across many areas of speech, in both formal and informal text. In the last century, numerous explanations have been given to describe what order people use for these binomials, from differences in semantics to differences in phonology. These rules describe primarily `frozen' binomials that exist in exactly one ordering and have lacked large-scale trials to determine efficacy.
  
  Online text provides a unique opportunity to study these lists in the context of informal text at a very large scale. In this work, we expand the view of binomials to include a large-scale analysis of both frozen and non-frozen binomials in a quantitative way. Using this data, we then demonstrate that most previously proposed rules are ineffective at predicting binomial ordering. By tracking the order of these binomials across time and communities we are able to establish additional, unexplored dimensions central to these predictions.
  
  Expanding beyond the question of individual binomials, we also explore the global structure of binomials in various communities, establishing a new model for these lists and analyzing this structure for non-frozen and frozen binomials. Additionally, novel analysis of trinomials --- lists of length three --- suggests that none of the binomials analysis applies in these cases. Finally, we demonstrate how large data sets gleaned from the web can be used in conjunction with older theories to expand and improve on old questions.
  
\end{abstract}


\maketitle

\section{Introduction}
Lists are extremely common in text and speech, and
the ordering of items in a list can often reveal information.
For instance, orderings can denote relative importance, such as on a to-do list,
or signal status, as is the case for author lists of scholarly publications.
In other cases, orderings might come from cultural or historical conventions.
For example, `red, white, and blue' is a specific ordering of colors that is recognizable to those familiar with American culture.

The orderings of lists in text and speech is a subject that has been repeatedly touched upon for more than a century.
By far the most frequently studied aspect of list ordering is the \textit{binomial}, a list of two words usually separated by a conjunction such as `and' or `or', which is the focus of our paper.
The academic treatment of binomial orderings dates back more than a century to Jespersen~\cite{jespersen_1905}, 
who proposed in 1905 that the ordering of many common English binomials could be predicted by the rhythm of the words. 
In the case of a binomial consisting of a monosyllable and a disyllable,
the prediction was that the monosyllable would appear first followed by the conjunction `and'.
The idea was that this would give a much more standard and familiar syllable stress to the overall phrase, e.g., the binomial 
`bread and butter' would have the preferable rhythm compared to `butter and bread.'

This type of analysis is meaningful when the two words in the binomial nearly always appear in the same ordering.
Binomials like this that appear in strictly one order (perhaps within the confines of some text corpus), are
commonly termed \emph{frozen binomials}~\cite{malkiel1959studies,gustafsson1976frequency}.
Examples of frozen binomials include `salt and pepper' and `pros and cons', and
explanations for their ordering in English and other languages have become increasingly complex.
Early work focused almost exclusively on common frozen binomials, often drawn from everyday speech. More recent work has expanded this view to include nearly frozen binomials, binomials from large data sets such as books, and binomials of particular types such as food, names, and descriptors \cite{cooperandross, benor2006chicken, mollin2012revisiting, wright2005ladies, motschenbacher2013gentlemen, hagerty2015ladies}. Additionally, explanations have increasingly focused on meaning rather than just sound, implying value systems inherent to the speaker or the culture of the language's speakers (one such example 
is that men are usually listed before women in English~\cite{fredandwilma}). 
The fact that purely phonetic explanations have been insufficient suggests that list orderings rely at least partially on semantics,
and it has previously been suggested that these semantics could be revealing about the culture in which the speech takes place~\cite{cooperandross}. Thus, it is possible that understanding these orderings could reveal biases or values held by the speaker. 

Overall, this prior research has largely been confined to pristine examples, often relying on small samples of lists to form conclusions.
Many early studies simply drew a small sample of what the author(s) considered some of the more representative or prominent binomials in whatever language they were studying~\cite{abraham1950coordinates, malkiel1959studies, sobkowiak1993unmarked, jespersen_1905, Scott_rhythm, fenk1989word, allan1987hierarchies, bolinger1962binomials, cooperandross}.
Other researchers have used books or news articles~\cite{gustafsson1976frequency, benor2006chicken}, or small samples from the Web (web search results and Google books)~\cite{mollin2012revisiting}.
Many of these have lacked a large-scale text corpus and have relied on a focused set of statistics about word orderings.

Thus, despite the long history of this line of inquiry, there is an opportunity to extend it significantly by 
examining a broad range of questions
about binomials coming from a large corpus of online text data produced organically by many people. 
Such an analysis could produce at least two types of benefits.
First, such a study could
help us learn about cultural phenomena embedded in word orderings and how they vary across communities and over time. 
Second, such an analysis could become a case study for the extension of theories developed at small scales in this domain to a much larger context. 

\xhdr{The present work: Binomials in large-scale online text}  In this work, we use data from large-scale Internet text corpora to study binomials at a massive scale, drawing on text created by millions of users. Our approach is more wholesale than prior work - we focus on all binomials of sufficient frequency, without first restricting to small samples of binomials that might be frozen. We draw our data from news publications, wine reviews, and Reddit, which in addition to large volume, also let us characterize binomials in new ways, and analyze differences in binomial orderings across communities and over time.
Furthermore, the subject matter on Reddit leads to many lists about people
and organizations that lets us study orderings of proper names --- a key setting for word ordering which has been difficult to study by other means.

We begin our analysis by introducing several new key measures for the study of binomials,
including a quantity we call \emph{asymmetry} that measures how frequently a given binomial appears
in some ordering.
By looking at the distribution of asymmetries across a wide range of binomials, we find
that most binomials are not frozen, barring a few strong exceptions. 
At the same time, there may still be an ordering preference. For example, `10 and 20' is not a frozen binomial;
instead, the binomial ordering `10 and 20' appears 60\% of the time and `20 and 10' appears 40\% of time.

We also address temporal and community structure in collections of binomials.
While it has been recognized that the orderings of  binomials may change over time or between communities~\cite{mollin2012revisiting, abraham1950coordinates, malkiel1959studies, fenk1989word, allan1987hierarchies, bolinger1962binomials}, there has been little analysis of this change. 
We develop new metrics for the \emph{agreement} of binomial orderings across communities and the \emph{movement} of binomial orderings over time.
Using subreddits as communities,
these metrics reveal variations in orderings, some of which suggest cultural change influencing language.
For example, in one community, we find that over a period of 10 years, 
the binomial `son and daughter' went from nearly frozen to appearing
in that order only 64\% of the time.

While these changes do happen, they are generally quite rare. 
Most binomials --- frozen or not --- are ordered in one way about the same percentage of the time, regardless of community or the year.
We develop a null model to determine how much variation in binomial orderings we might expect
across communities and across time, if binomial orderings were randomly ordered according to
global asymmetry values. We find that there is less variation across time and communities in the data
compared to this model, implying that binomial orderings are indeed remarkably stable.

Given this stability, one might expect that the dominant ordinality of a given binomial
is still predictable, even if the binomial is not frozen. For example, one might expect
that the global frequency of a single word or the number of syllables in a word would
predict ordering in many cases. However, we find that these simple predictors are quite poor at determining binomial ordering.

On the other hand, we find that a notion of `proximity' is robust at predicting ordering in
some cases. Here, the idea is that the person producing the text will list the word that
is conceptually ``closer'' to them first --- a phenomenon related to a ``Me First'' principle of binomial orderings
suggested by Cooper and Ross~\cite{cooperandross}. One way in which we study this notion of proximity
is through sports team subreddits. For example, we find that when two NBA team names form a binomial
on a specific team's subreddit, the team that is the subject of the subreddit tends to appear first. 

The other source of improved predictions comes from using word embeddings \cite{mikolov2013distributed}: we find that a model based on the positions of words in a standard pre-trained word embedding can be a remarkably reliable predictor of binomial orderings. While not applicable to all words, such as names, this type of model is strongly predictive in most cases.

Since binomial orderings are in general difficult to predict individually, we explore a new way
of representing the global binomial ordering structure,
we form a directed graph where an edge from $i$ to $j$ means that $i$ tends to come before $j$ in binomials. 
These graphs show tendencies across the English language and also reveal peculiarities in the language of particular communities. For instance, in a graph formed from the binomials in a sports community, the names of sports teams and cities are closely clustered, showing that they are often used together in binomials. Similarly, we identify clusters of names, numbers, and years. 
The presence of cycles in these graphs are also informative. 
For example, cycles are rare in graphs formed from proper names in politics, suggesting a possible hierarchy of names,
and at the same time very common for other binomials. 
This suggests that no such hierarchy exists for most of the English language, 
further complicating attempts to predict binomial order.

Finally, we expand our work to include multinomials, which are lists of more than two words. 
There already appears to be more structure in trinomials (lists of three) compared to binomials. 
Trinomials are likely to appear in exactly one order, and when they appear in more than one order the last word is almost always the same across all instances. For instance, in one section of our Reddit data, `Fraud, Waste, and Abuse' appears 34 times, and  `Waste, Fraud, and Abuse' appears  20 times.
This could point to, for example, recency principles being more important in lists of three than in lists of two.  While multinomials were in principle part of the scope of past research in this area, they were difficult to study in smaller corpora, suggesting another benefit of working at our current scale.

\subsection{Related Work}

Interest in list orderings spans the last century~\cite{abraham1950coordinates,malkiel1959studies}, 
with a focus almost exclusively on binomials.
This research has primarily investigated frozen binomials, also called {\em irreversible binomials}, {\em fixed coordinates}, and {\em fixed conjuncts}~\cite{sobkowiak1993unmarked}, although some work has also looked at non-coordinate freezes where the individual words are nonsensical by themselves (e.g., `dribs and drabs')~\cite{sobkowiak1993unmarked}. 
One study has directly addressed \textit{mostly} frozen binomials~\cite{mollin2012revisiting}, and we expand the scope 
of this paper by exploring the general question of how frequently binomials appear in a particular order.
Early research investigated languages other than English~\cite{malkiel1959studies,abraham1950coordinates}, but
most recent research has worked almost exclusively with English. 
Overall, this prior research can be separated into three basic categories --- phonological rules, semantic rules, and metadata rules.

\xhdr{Phonology}
The earliest research on binomial orderings proposed mostly phonological explanations, particularly rhythm~\cite{jespersen_1905, Scott_rhythm}. 
Another highly supported proposal is Panini's Law, which claims that words with fewer syllables come first~\cite{pinker1979speakers}; we find only very mild preference for this type of ordering.
Cooper and Ross's work expands these to a large list of rules, many overlapping, and suggests that they can compound~\cite{cooperandross}; a number of subsequent papers have expanded on their work~\cite{sobkowiak1993unmarked, bolinger1962binomials, fredandwilma, pinker1979speakers}.

\xhdr{Semantics}
There have also been a number of semantic explanations, mostly in the form of categorical tendencies (such as `desirable before undesirable') that may have cultural differences~\cite{abraham1950coordinates, malkiel1959studies}. The most influential of these may be the `Me First' principle codified by Cooper and Ross. This suggests that the first word of a binomial tends to follow a hierarchy that favors `here', `now', present generation, adult, male, and positive. Additional hierarchies also include a hierarchy of food, plants vs.\ animals, etc.~\cite{cooperandross}.

\xhdr{Frequency}
More recently, it has been proposed that the more cognitively accessible word might come first, which often means the word the author sees or uses most frequently~\cite{mcdonald1993word}. There has also been debate on whether frequency may encompass most phonological and semantic rules that have been previously proposed~\cite{fenk1989word, benor2006chicken}. 
We find that frequency is in general a poor predictor of word ordering.

\xhdr{Combinations}
Given the number of theories, there have also been attempts to give a hierarchy of rules and study their interactions~\cite{benor2006chicken,mollin2012revisiting}. This research has complemented the
proposals of Cooper and Ross~\cite{cooperandross}. 
These types of hierarchies are also presented as explanations for the likelihood of a binomial becoming frozen~\cite{mollin2012revisiting}.

\xhdr{Names}
Work on the orderings of names has been dominated by a single phenomenon: men's names usually come before women's names. Explanations range from a power differential, to men being more `agentic' within `Me First', to men's names being more common or even exhibiting more of the phonological features of words that usually come first~\cite{hagerty2015ladies, mollin2012revisiting, mcdonald1993word, cooperandross, fenk1989word, fredandwilma, hegarty2011gentlemen, wright2005ladies}. However, it has also been demonstrated that this preference may be affected by the author's own gender and relationship with the people named~\cite{wright2005ladies, hegarty2011gentlemen}, as well as context more generally~\cite{kesebir2017word}.

\xhdr{Orderings on the Web}
List orderings have also been explored in other Web data,
specifically on the ordering of tags applied to images~\cite{nwana2016ordered}. There is evidence that these tags are ordered intentionally by users, and that a bias to order tag A before tag B may be influenced by historical precedent in that environment but also by the relative importance of A and B~\cite{nwana2016ordered}. 
Further work also demonstrates that exploiting the order of tags on images can improve models that rank those images~\cite{nwana2017quote}. 

\section{Data}
We take our data mostly from Reddit, a large social media website divided into subcommunities called `subreddits' or `subs'.
Each subreddit has a theme (usually clearly expressed in its name), and we have focused our study on subreddits primarily in sports and politics, in part because of the richness of proper names in these domains: r/nba, r/nfl, r/politics, r/Conservative, r/Libertarian, r/The\_Donald, r/food, along with a variety of NBA team subreddits (e.g., r/rockets for the Houston Rockets).
Apart from the team-specific and food subreddits, these are among 
the largest and most heavily used subreddits~\cite{metrics}. 
We gather text data from comments made by users in discussion threads.
In all cases, we have data from when the subreddit started until mid-2018.
(Data was contributed by Cristian Danescu-Niculescu-Mizil.)
Reddit in general, and the subreddits we examined in particular, are rapidly growing,
both in terms of number of users and number of comments. 

\begin{figure}[tb]
    \includegraphics[width=.35\textwidth]{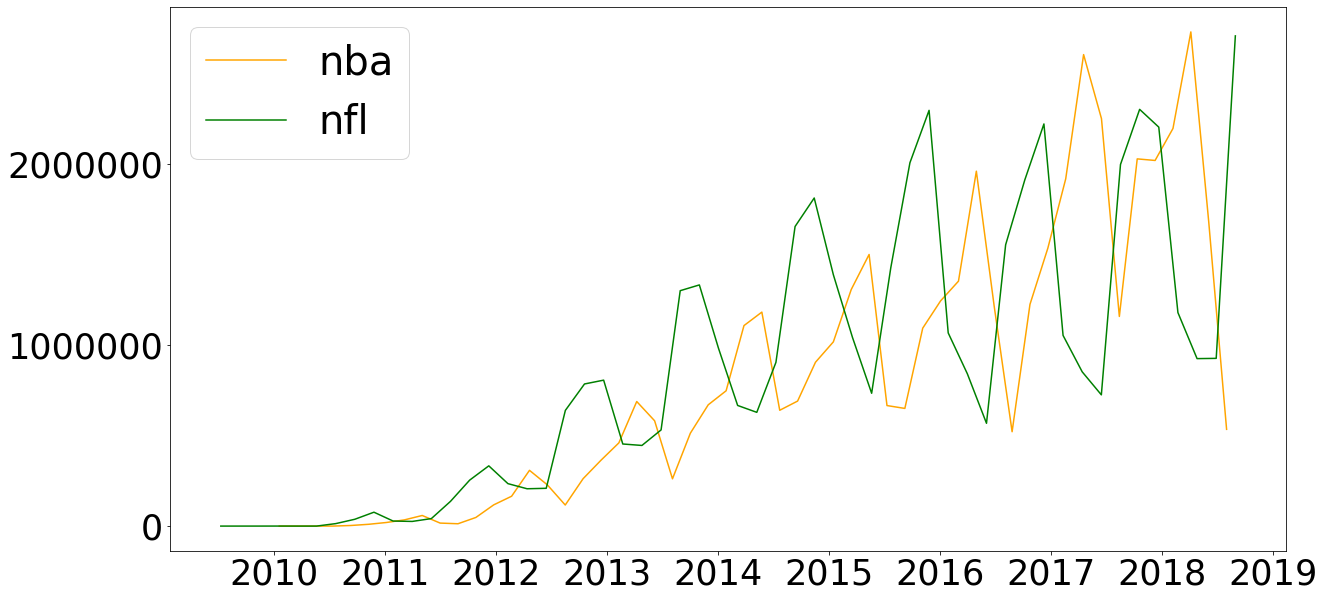}
     
  \caption{Histogram of comment timestamps for r/nba and r/nfl. 
  Both subreddits exhibit a seasonal structure. 
  The number of comments is increasing for all subreddits.}
\label{fig:comment_timestamps}    
\end{figure}

Some of the subreddits we looked at (particularly sports subreddits) exhibited very distinctive `seasons', where commenting spikes (Fig.~\ref{fig:comment_timestamps}).
These align with, e.g., the season of the given sport.
When studying data across time, our convention is to bin the data by year,
but we adjust the starting point of a year based on these seasons.
Specifically, a year starts in May for r/nfl, August for r/nba, and February for all politics subreddits.

\begin{table*}[t]
\caption{Summary statistics of subreddit list data that we investigate in this paper.}
  \label{tab:data summary}
    \begin{tabular}{lllllll}
    \toprule
                   & Total Comments & Seasons in Operation & Total Lists & Name Lists &  Unique Names\\
    \midrule
    r/food         & 4580723                  & 10                  & 924457                & -                    & -  \\
    r/nba          & 44661753                 & 10                  & 4807623               & 1826090              & 29336 \\
    r/nfl          & 46515339                 & 9                   & 5855011               & 1287597              & 35014\\
    r/Conservative & 2772010                  & 11                  & 859252                & 77593                & 6742\\
    r/Libertarian  & 5485366                  & 11                  & 2318685               & 146729               & 8964\\
    r/politics     & 86208050                 & 12                  & 26142750              & 1854730              & 43316\\
    r/The\_Donald  & 2576974                  & 3                   & 3964936               & 388339               & 18891\\
    \bottomrule
    \end{tabular}
\end{table*}

We use two methods to identify lists from user comments:  `All Words' and `Names Only',
with the latter focusing on proper names. In both cases, we collect a number of lists
and discard lists for any pair of words that appear fewer than 30 times within the time
frame that we examined (see Table~\ref{tab:data summary} for summary statistics).

The All Words method simply searches for two words $A$ and $B$ separated by `and' or `or', where a word is merely a series of characters separated by a space or punctuation. This process only captures lists of length two, or binomials.  We then filter out lists containing words from a collection of stop-words that, by their grammatical role or formatting structure, are almost exclusively involved in false positive lists.  
No metadata is captured for these lists beyond the month and year of posting.

\begin{figure}[tb]
    \includegraphics[width=.30\textwidth]{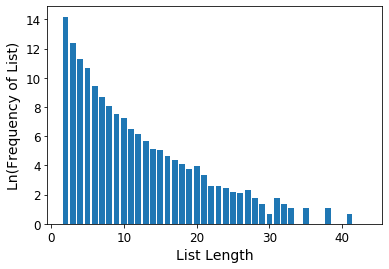}
  \caption{A histogram of the log frequency of lists of various lengths, where we use name lists for r/nba. 
  In this case, there is no filtering applied, but we cap list length at 50.}
   
  \label{fig:log_frequency}    
\end{figure}

The Names Only method uses a curated list of full names relevant to the subreddit,
focusing on sports and politics.
For sports, we collected names of all NBA and NFL player
active during 1980--2019 from \textit{basketball-reference.com} and \textit{pro-football-reference.com}.
For politics, we collected the names of congresspeople from \textit{the @unitedstates project}~\cite{politics}.
To form lists, we search for any combination of any part of these names such that at least two partial names are separated by 
`and', `or', `v.s.', `vs', or `/' and the rest are separated by `,'. 
While we included a variety of separators, about 83\% of lists include only `and', about 17\% include `or' and the rest of the separators are negligible.
Most lists that we retrieve in this way are of length 2, but we also found lists up to length 40 (Fig.~\ref{fig:log_frequency}).
Finally, we also captured full metadata for these lists,
 including a timestamp, the user, any flairs attributed to the user (short custom text that appears next to the username), and other information.
 
 We additionally used wine reviews and a variety of news paper articles for additional analysis. The wine data gives reviews of wine from WineEnthusiast and is hosted on Kaggle \cite{wine}. While not specifically dated, the reviews were scraped between June and November of 2017. There are 20 different reviewers included, but the amount of reviews each has ranges from tens to thousands. The news data consists of news articles pulled from a variety of sources, including (in random order) the New York Times, Breitbart, CNN, the Atlantic, Buzzfeed News, National Review, New York Post, NPR, Reuters, and the Washington Post. The articles are primarily from 2016 and early 2017 with a few from 2015. The articles are scraped from home-page headline and RSS feeds \cite{news}. Metadata was limited for both of these data sets. \footnote{Example code to identify binomials in Reddit data can be found at https://www.cornell.edu/~kvank/}

%
%

\section{Dimensions of Binomials}
In this paper we introduce a new framework to interpret binomials,
based on three properties: 
\emph{asymmetry} (how frozen a binomial is),
\emph{movement} (how binomial orderings change over time), and
\emph{agreement} (how consistent binomial orderings are between communities), which
we will visualize as a cube with three dimensions.
Again, prior work has focused essentially entirely on asymmetry, and we argue that this can only really
be understood in the context of the other two dimensions.

For this paper we will use the convention \{A,B\} to refer to an unordered pair of words, and [A,B] to refer to an ordered pair where A comes before B. We say that [A,B] and [B,A] are the two possible {\em orientations} of \{A,B\}. 

\subsection{Definitions}
Previous work has one main measure of binomials --- their `frozen-ness'. A binomial is `frozen' if it always appears with a particular order. 
For example, if the pair \{`arrow', `bow'\} always occurs as [`bow', `arrow'] and never as [`arrow', `bow'], then it is frozen.
This leaves open the question of how describe the large number of binomials that are not frozen.
To address this point, we instead consider the {\em ordinality} of a list, or how often the list is `in order' according to some arbitrary underlying reference order. 
Unless otherwise specified, the underlying order is assumed to be alphabetical. 
If the list [`cat', `dog'] appears 40 times and the list [`dog', `cat'] 10 times, then the list \{`cat', `dog'\} would have an ordinality of 0.8.

Let $n_{x,y}$ be the number of times the ordered list $[x,y]$ appears, and
let $f_{x,y} = n_{x,y} / (n_{x,y} + n_{y,x})$ be the fraction of times that the unordered version of the list appears in that order.
We formalize ordinality as follows.
\theoremstyle{definition}
\begin{definition}[Ordinality]
Given an ordering $<$ on words (by default, we assume alphabetical ordering),
the \textbf{ordinality} $o_{x,y}$ of the pair $\{x,y\}$ is
equal to $f_{x,y}$ if $x < y$ and $f_{y,x}$ otherwise.
\end{definition}

Similarly, we introduce the concept of asymmetry in the context of binomials, which
is how often the word appears in its dominant order.
In our framework, a `frozen' list is one with ordinality 0 or 1 and
would be considered a high asymmetry list, with asymmetry of 1. 
A list that appears as [`A', `B'] half of the time and [`B', `A'] half of the time (or with ordinality 0.5) 
would be considered a low asymmetry list, with asymmetry of 0.

\begin{definition}[Asymmetry]
The \textbf{asymmetry} of an unordered list $\{x,y\}$ is 
$A_{x,y} = 2 \cdot \lvert o_{x,y} - 0.5 \rvert$.
\end{definition}

The Reddit data described above gives us access to new dimensions of binomials not previously addressed. 
We define movement as how the ordinality of a list changes over time
\begin{definition}[Movement]
Let $o_{x,y,t}$ be the ordinality of an unordered list $\{x,y\}$ for data in year $t \in T$.
The \textbf{movement} of $\{x,y\}$ is $M_{x,y} = \max_{t \in T} o_{x,y,t} - \min_{t \in T} o_{x,y,t}$.
\end{definition}
And agreement describes how the ordinality of a list differs between different communities.
\begin{definition}[Agreement]
Let $o_{x,y,c}$ be the ordinality of an unordered list ${x,y}$ for data in community (subreddit) $c \in C$.
The \textbf{agreement} of $\{x,y\}$ is $A_{x,y} = 1 - (\max_{c \in C} o_{x,y,c} - \min_{c \in C} o_{x,y,c})$.
\end{definition}

\subsection{Dimensions}

\begin{figure}
    \includegraphics[width=.45\textwidth]{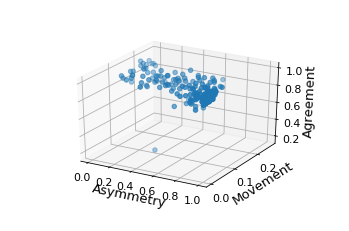}
    \caption{309 binomials that occur at least 30 times per year in r/politics, r/nba, and r/nfl mapped on to the 3-dimensional cube. The point on the bottom left is \{`10', `20'\}.}
     
    \label{fig:cube}
\end{figure}

Let the point $(A,M,G)_{x,y}$ be a vector of the asymmetry, movement, and agreement for some unordered list $\{x,y\}$. These vectors then define a 3-dimensional space in which each list occupies a point. Since our measures for asymmetry, agreement, and movement are all defined from 0 to 1, their domains form a unit cube (Fig.~\ref{fig:cube}). 
The \textit{corners} of this cube correspond to points with coordinates are entirely made up of 0s or 1s. 
By examining points near the corners of this cube, we can get a better understanding of the range of binomials. 
Some corners are natural --- it is easy to imagine a high asymmetry, low movement, high agreement binomial --- such as \{`arrow', `bow'\} from earlier. 
On the other hand, we have found no good examples of a high asymmetry, low movement, low agreement binomial. There are a few unusual examples, such as \{10, 20\}, which has 0.4 asymmetry, 0.2 movement, and 0.1 agreement and is clearly visible as an isolated point in Fig.~\ref{fig:cube}.

{}

\omt{
\begin{table*}[htbp]
  \centering
  \caption{Points closest to cube corners Fig.~\ref{fig:cube}. 
  The tuples represent the dimensions of the list (asymmetry, movement, agreement).}
  \label{tab: dimensional corner examples}
  \begin{tabular}{*{11}{l}}
    \toprule
    & \multicolumn{2}{c}{High Agreement} & \multicolumn{2}{c}{Low Agreement} \\
    \cmidrule(lr){2-3}
    \cmidrule(lr){4-5}
    & Low Movement & High Movement & Low Movement & High Movement \\
    \midrule
    High Asymmetry & \{bow, arrow\}   & \{10, 20\}      & \{name, number\} & \{10, 20\}\\
     &(0.0, 0.0, 1.0) &(0.6, 0.8, 0.9)&(0.4, 0.2, 0.7) & (0.6, 0.8, 0.9) \\
    Low Asymmetry  &\{chicago, detroit\}&{10, 20}       &\{school, work\}&\{10, 20\}\\
     &(0.9, 0.1, 0.9)   &(0.6, 0.8, 0.9)& (2.0, 0.1, 0.8)&(0.6, 0.8, 0.9)\\
    \bottomrule
  \end{tabular}
  \label{tab:case-studies}
\end{table*}
}

\begin{figure}[h!]
    \includegraphics[width=.40\textwidth]{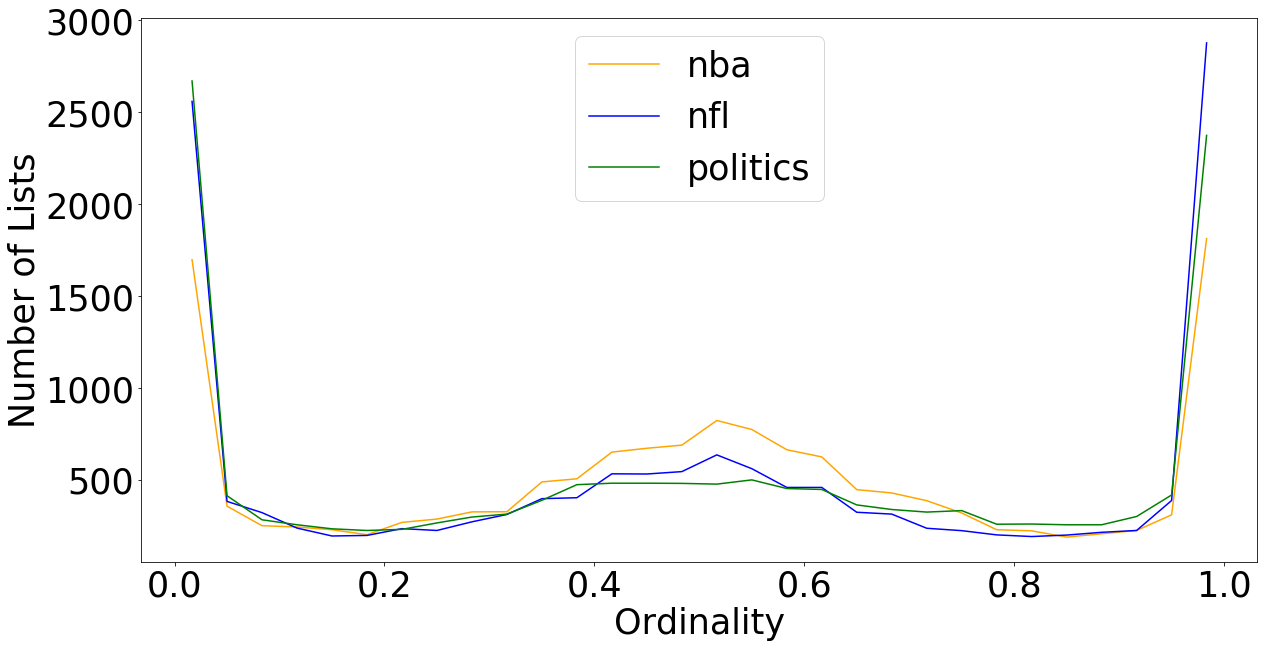}
  \caption{Histograms of the alphabetical orientation of the 14920 most common binomials within r/nba, r/nfl and r/politics. Note that while there are many frozen binomials (with orientation of 0 or 1), the rest of the binomials appear to be roughly normally distributed around 0.5.
  }
   
  \label{fig: binomial percs plots}
\end{figure}

\xhdr{Asymmetry}
While a majority of binomials have low asymmetry, almost all previous work has focused exclusively on high-asymmetry binomials. In fact, asymmetry is roughly normally distributed across binomials with an additional increase of highly asymmetric binomials (Fig.~\ref{fig: binomial percs plots}). This implies that previous work has overlooked the vast majority of binomials, and an investigation into whether rules proposed for highly asymmetric binomials also functions for other binomials is a core piece of our analysis.

\xhdr{Movement}
The vast majority of binomials have low movement.
However, the exceptions to this can be very informative. 
Within r/nba a few of these pairs show clear change in linguistics and/or culture. 
The binomial [`rpm', `vorp'] (a pair of basketball statistics) started at 0.74 ordinality and within three years dropped to 0.32 ordinality, showing a potential change in users' representation of how these statistics relate to each other. 
In r/politics, [`daughter', `son'] moved from 0.07 ordinality to 0.36 ordinality over ten years. This may represent a cultural shift in how users refer to children, or a shift in topics discussed relating to children.
And in r/politics, ['dems', 'obama'] went from 0.75 ordinality to 0.43 ordinality from 2009--2018, potentially
reflecting changes in Obama's role as a defining feature of the Democratic Party. Meanwhile the ratio of unigram frequency of `dems' to `obama' actually increased from 10\% to 20\% from 2010 to 2017. Similarly, [`fdr', `lincoln'] moved from 0.49 ordinality to 0.17 ordinality from 2015--2018. 
This is particularly interesting, since in 2016 `fdr' had a unigram frequency 20\% higher than `lincoln', but in 2017 they are almost the same. This suggests that movement could be unrelated to unigram frequency changes. Note also that the covariance for movement across subreddits is quite low \ref{tab:movement covariance}, and movement in one subreddit is not necessarily reflected by movement in another.

\begin{table}[]
    \caption{Covariance table for movement in r/nba, r/nfl and r/politics.}
    \label{tab:movement covariance}
    \begin{tabular}{cccc}
        \toprule
          &  r/nba & r/nfl & r/politics\\
          \midrule
            r/nba & 0.00426 & 0.00240 & 0.00260 \\
            r/nfl & 0.00240 & 0.00468 & 0.00320 \\
            r/politics & 0.00560 & 0.00320 & 0.00579 \\
            \bottomrule
    \end{tabular}
\end{table}

\xhdr{Agreement}
Most binomials have high agreement (Table~\ref{tab:asymmetry comps}) but again the counterexamples are informative. 
For instance, [`score', `kick'] has ordinality of 0.921 in r/nba and 0.204 in r/nfl. This likely points to the fact that American football includes field goals. A less obvious example is the list [`ceiling', `floor']. 
In r/nba and r/nfl, it has ordinality 0.44, and in r/politics, 
it has ordinality 0.27.

There are also differences among proper nouns.
One example is [`france', `israel'], which has ordinality
0.6 in r/politics,
0.16 in r/Libertarian, and
0.51 in r/The\_Donald (and the list does not appear in r/Conservative).
And the list [`romney', `trump'] has ordinality
0.48 in r/poltics,
0.55 in r/The\_Donald, and
0.73 in r/Conservative.


\begin{table*}[]
\caption{The average difference in asymmetry between the same binomial in various subreddits. The difference between r/nba and r/nfl is 0.062.}
\label{tab:asymmetry comps}
\begin{tabular}{lllll}
\toprule
               & r/Conservative & r/Libertarian & r/politics & r/the\_Donald \\
\midrule
               
r/Conservative &                & 0.057         & 0.049      & 0.055         \\
r/Libertarian  & 0.057          &               & 0.053      & 0.059         \\
r/politics     & 0.049          & 0.053         &            & 0.052         \\
r/The\_Donald  & 0.055          & 0.059         & 0.052      &              \\
\bottomrule
\end{tabular}
\end{table*}

\section{Models And Predictions}
In this section, we establish a null model 
under which different communities or time slices have the same probability
of ordering a binomial in a particular way. With this, we would expect to see variation
in binomial asymmetry. We find that our data shows smaller variation
than this null model predicts, suggesting that binomial orderings
are extremely stable across communities and time.
From this, we might also expect that orderings are predictable; but we find
that standard predictors in fact have limited success.

\subsection{Stability of Asymmetry}
Recall that the asymmetry of binomials with respect to alphabetic order (excluding frozen binomials) 
is roughly normal centered around $0.5$ (Fig.~\ref{fig: binomial percs plots}).
One way of seeing this type of distribution would be if binomials are ordered randomly, with $p=0.5$ for each order.
In this case, if each instance $l$ of a binomial $\{x,y\}$ takes value 0 (non-alphabetical ordering) or 1 (alphabetical ordering), 
then $l \sim \text{Bernoulli}(0.5)$. 
If $\{x,y\}$ appears $n$ times, then the number of instances of value 1 is
distributed by $W \sim \text{Bin}(n, 0.5)$, and $W / n$
is approximately normally distributed with mean 0.5.

One way to test this behavior is to first estimate $p$ for each list within each community.
If the differences in these estimates are not normal, then the above model is incorrect.
We first omit frozen binomials before any analysis.
Let $L$ be a set of unordered lists and $C$ be a set of communities.
We estimate $p$ for list $l \in L$ in community $c \in C$ by 
$\hat{p}_{l,c} = o_{l,c}$, the ordinality of $l$ in $C$.
Next, for all $l \in L$ let $p^*_{l} = \max_{c \in C}(\hat{p}_{l, c}) - \min_{ c \in C}(\hat{p}_{l, c})$.
The distribution of $p^*_{l}$ over $l \in L$ has median 0, mean 0.0145, and standard deviation 0.0344. 
We can perform a similar analysis over time.
Define $Y$ as our set of years, and $\hat{p}_{l, y} = o_{l,y}$ for $y \in Y$ our estimates.
The distribution of $p'_{l} = \max_{y \in Y}(\hat{p}_{l, y}) - \min_{y \in Y}(\hat{p}_{l, y})$ over $l \in L$
has median 0.0216, mean 0.0685, and standard deviation 0.0856. 
The fact that $p$ varies very little across both time and communities suggests that there is
some $p_l$ for each $l \in L$ that is consistent across time and communities, which
is not the case in the null model, where these values would be normally distributed.

\begin{figure}[h!]
    \includegraphics[width=.35\textwidth]{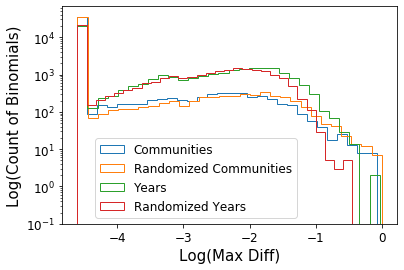}
    \caption{Histogram of the maximum difference in $p_l$ for all lists $l$ across communities and years, on a log-log scale. 
    We add 0.01 to all differences to show cases with a difference of 0, which is represented as the bar on the left of the graph (mostly due to frozen binomials). 
    We sampled 40000 instances for this graph, since there was variation in the number of binomials across years and communities.}  
     
    \label{fig: max diff variability}
\end{figure}

We also used a bootstrapping technique to understand the mean variance in ordinality for lists over communities and years.
Specifically, let $o_{l, c, y}$ be the ordinality of list  $l$ in community $c$ and year $y$,
$O_l$ be the set of $o_{l,c,y}$ for a given list $l$, and $s_l$ be the standard deviation of $O_l$.
Finally, let $\bar{s}$ be the average of the $s_l$.
We re-sample data by randomizing the order of each binomial instance, 
sampling its orderings by a binomial random variable with success probability equal to its ordinality across all seasons and communities ($p_l$).
We repeated this process to get samples estimates $\{\bar{s}_1, \ldots, \bar{s}_{k}\}$, where $k$ is the size of the set of seasons and communities.
These averages range from 0.0277 to 0.0278 and are approximately normally distributed
(each is a mean over an approximately normal scaled Binomial random variable).
However, $\bar{s} = 0.0253$ for our non-randomized data. 
This is significantly smaller than the randomized data and implies that the true variation in $p_l$ across time and communities is even smaller than a binomial distribution would predict.
One possible explanation for this is that each instance of $l$ is not actually independent, but is in fact anti-correlated, violating one of the conditions of the binomial distribution. An explanation for that could be that users attempt to draw attention by intentionally going against the typical ordering~\cite{malkiel1959studies}, but it is an open question what the true model is and why the variation is so low. Regardless, it is clear that the orientation of binomials varies very little across years and communities (Fig.~\ref{fig: max diff variability}).

\subsection{Prediction Results}
Given the stability of binomials within our data, we now try to predict their ordering.
We consider deterministic or rule-based methods that predict the order for a given binomial.
We use two classes of evaluation measures for success on this task:
(i) by token --- judging each instance of a binomial separately; and
(ii) by type --- judging all instances of a particular binomial together. 
We further characterize these into weighted and unweighted.

To formalize these notions, first consider any unordered list $\{x,y\}$
that appears $n_{x,y}$ times in the orientation $[x,y]$ and $n_{y,x}$ times in the orientation $[y,x]$.
Since we can only guess one order, we will have either $n_{x,y}$ or $n_{y,x}$ successful
guesses for $\{x,y\}$ when guessing by token.
The unweighted token score (UO) and weighted token score (WO) are the macro and micro averages of this accuracy.

If predicting by type, let $S$ be the lists such that the by-token prediction is successful at least half of the time.
Then the unweighted type score (UT) and weighted type score (WT) are the macro and micro averages of $S$.

\xhdr{Basic Features}
We first use predictors based on rules that have previously been proposed in the literature:
word length, number of phonemes, number of syllables, alphabetical order, and frequency. 
We collect all binomials but make predictions only on binomials appearing at least 30 times total, stratified by subreddit. However, none of these features appear to be particularly predictive across the board (Table~\ref{tab:prediction_results}). A simple linear regression model predicts close to random, which bolsters the evidence that these classical rules for frozen binomials are not predictive for general binomials.

\begin{table}[]
\caption{Accuracy of binomial orientation predictions using a number of basic rules. The scoring was done based on ``unweighted type'' scoring, and statistics are given based on the scores across the subreddits.}
\label{tab:prediction_results}

\begin{tabular}{@{}llll@{}}
\toprule
                  & Min  & Mean  & Max  \\ \midrule
Alphabetical      & 0.365 & 0.462  & 0.512 \\
Length            & 0.483 & 0.563  & 0.639 \\
Phonemes          & 0.477 & 0.561  & 0.642 \\
Syllables         & 0.588 & 0.654  & 0.714 \\
Unigram Frequency & 0.553 & 0.599  & 0.629 \\ \bottomrule
\end{tabular}
\end{table}


Perhaps the oldest suggestion to explain binomial orderings is that if there are two words A and B, and A is monosyllabic and B is disyllabic, then A comes before B~\cite{jespersen_1905}. Within r/politics, we gathered an estimate of number of syllables for each word as given by a variation on the CMU Pronouncing Dictionary~\cite{bartlett-etal-2009-syllabification} 
(Tables~\ref{tab: Syllable Table}~and~\ref{tab:syllable_frac}).
In a weak sense, Jespersen was correct that monosyllabic words come before disyllabic words more often than not; and more generally, shorter words come before longer words more often than not. However, as predictors, these principles are close to random guessing.

\begin{table}[]
\centering
\caption{Count for number of syllables in first and second word of all binomials in r/politics. First word is rows, second word is columns. Overall, shorter words are significantly more likely to come before longer words (see also Table~\ref{tab:syllable_frac}).}
\label{tab: Syllable Table}
\begin{tabular}{@{}l|lllll@{}}
  & 1     & 2     & 3     & 4    & 5    \\ \midrule
1 & 24388 & 17964 & 8707  & 3269 & 719  \\
2 & 17354 & 18296 & 10507 & 4364 & 1139 \\
3 & 7758  & 8462  & 7271  & 3929 & 1134 \\
4 & 3148  & 3313  & 2889  & 1975 & 721  \\
5 & 500   & 621   & 642   & 472  & 197 
\end{tabular}
\end{table}

\begin{table}[]
\caption{Similar to Table~\ref{tab: Syllable Table}, this table displays the fraction of of lists with this ordering of that size. So, [2,1] is 0.491 and [1,2] is 0.509, meaning monosyllables slightly tend to come before disyllables. This is cut off at 5 syllables, as more syllables has too few instances.}
\label{tab:syllable_frac}
\begin{tabular}{l|llllllll}
  & 1    & 2    & 3    & 4    & 5      \\
  \toprule
1 & 1.0    & 0.509 & 0.529 & 0.509 & 0.590   \\
2 & 0.491 & 1.0    & 0.554 & 0.568 & 0.647    \\
3 & 0.471 & 0.446 & 1.0    & 0.576 & 0.639  \\
4 & 0.491 & 0.432 & 0.424 & 1.0    & 0.604  \\
5 & 0.410 & 0.353 & 0.361 & 0.396 & 1.0     \\

\end{tabular}
\end{table}

\xhdr{Paired Predictions}
Another measure of predictive power is predicting which of two binomials has higher asymmetry. 
In this case, we take two binomials with very different asymmetry and try to predict which has higher asymmetry 
by our measures (we use the top-1000 and bottom-1000 binomials in terms of asymmetry for these tasks). For instance, we may predict that [`red', `turquoise'] is more asymmetric than [`red', `blue'] because the differences in lengths is more extreme. Overall, the basic predictors from the literature are not very successful (Table~\ref{tab: paired results}).
\begin{table*}[]
\caption{Paired prediction results.}
\label{tab: paired results}
    \begin{tabular}{l @{\qquad} llll  @{\qquad} ll @{\qquad} l}
        \toprule
         predictor &  r/Conservative & r/Libertarian & r/politics & r/The\_Donald & r/nba & r/nfl & r/food\\
         \midrule
         Alphabetical & 0.501 & 0.417 & 0.539 & 0.631 & 0.360 & 0.351 & 0.234 \\
         Length & 0.445 & 0.481 & 0.330 & 0.505 & 0.482 & 0.324 & 0.447\\
         Phonemes & 0.430 & 0.460 & 0.324 & 0.513 & 0.428 & 0.447 & 0.417\\
         Syllables & 0.357 & 0.439 & 0.505 & 0.328 & 0.477 & 0.257 & 0.453\\
         \bottomrule
    \end{tabular}
    \label{tab:paired prediction}
\end{table*}

\begin{table}[]
\centering
\caption{The accuracy using "unweighted type" for only frozen binomials, here defined as binomials with asymmetry above 0.97. The results suggest that these rules are equally ineffective for frozen and non-frozen binomials.}
\label{tab:frozen_results}
\begin{tabular}{@{}llll@{}}
\toprule
             & Min & Mean & Max \\
             \midrule
Alphabetical & 0.46 & 0.51  & 0.55 \\
Length       & 0.37 & 0.45  & 0.54 \\
Phonemes     & 0.33 & 0.44  & 0.51 \\
Sylables     & 0.39 & 0.45  & 0.55 \\
Unigrams     & 0.34 & 0.47  & 0.58 \\
\bottomrule
\end{tabular}
\end{table}

\xhdr{Word Embeddings}
If we turn to more modern approaches to text analysis, one of the most common is {\em word embeddings} \cite{mikolov2013distributed}. Word embeddings assign a vector $x_i$ to each word $i$ in the corpus, such that the relative position of these vectors in space encode information lingustically relevant relationships among the words. Using the Google News word embeddings, via a simple logistic model, we produce a vector $v^*$ and predict the ordering of a binomial on words $i$ and $j$ from $v^* \cdot(x_i - x_j)$.  In this sense, $v^*$ can be thought of as a ``sweep-line'' direction through the space containing the word vectors, such that the ordering along this sweep-line is the predicted ordering of all binomials in the corpus. This yields surprisingly accurate results, with accuracy ranging from 70\% to 85\% across various subreddits (Table~\ref{tab:embeddings}), and 80-100\% accuracy on frozen binomials. This is by far the best prediction method we tested. It is important to note that not all words in our binomials could be associated with an embedding, so it was necessary to remove binomials containing words such as names or slang. However, retesting our basic features on this data set did not show any improvement, implying that the drastic change in predictive power is not due to the changed data set.

\begin{table}[]
\caption{Results of logistic regression based on word embeddings. This is by far our most successful model. Note that not all words in our binomials were found in the word embeddings, leaving about 70--97\% of the binomials usable.}
\label{tab:embeddings}
\begin{tabular}{@{}lll@{}}
\toprule
             & Accuracy \\
\midrule
r/nba          & 0.70  \\
r/nfl          & 0.75   \\
r/Conservative & 0.83    \\
r/Libertarian  & 0.87    \\
r/politics     & 0.74    \\
r/The\_Donald  & 0.70     \\
r/food         & 0.85     \\
\bottomrule
\end{tabular}
\end{table}

\section{Proper Nouns and the Proximity Principle}
Proper nouns, and names in particular, have been a focus within the literature on frozen binomials~\cite{hagerty2015ladies, mollin2012revisiting, mcdonald1993word, cooperandross, fenk1989word, fredandwilma, wright2005ladies, hegarty2011gentlemen, kesebir2017word, kinpairs}, but these studies have largely concentrated on the effect of gender in ordering~\cite{hagerty2015ladies, mollin2012revisiting, mcdonald1993word, cooperandross, fenk1989word, fredandwilma, wright2005ladies, hegarty2011gentlemen, kesebir2017word}. 
With Reddit data, however, we have many conversations about large numbers of celebrities, with significant background information on each. 
As such, we can investigate proper nouns in three subreddits: r/nba, r/nfl, and r/politics. 
The names we used are from NBA and NFL players (1970--2019) and congresspeople (pre-1800 and 2000--2019) respectively. 
We also investigated names of entities for which users might feel a strong sense of identification, such as 
a team or political group they support, or a subreddit to which they subscribe. 
We hypothesized that the group with which the user identifies the most would come first in binomial orderings. 
Inspired by the `Me First Principle', we call this the {\em Proximity Principle}.

\begin{figure}
    \centering
    \includegraphics[width=.40\textwidth]{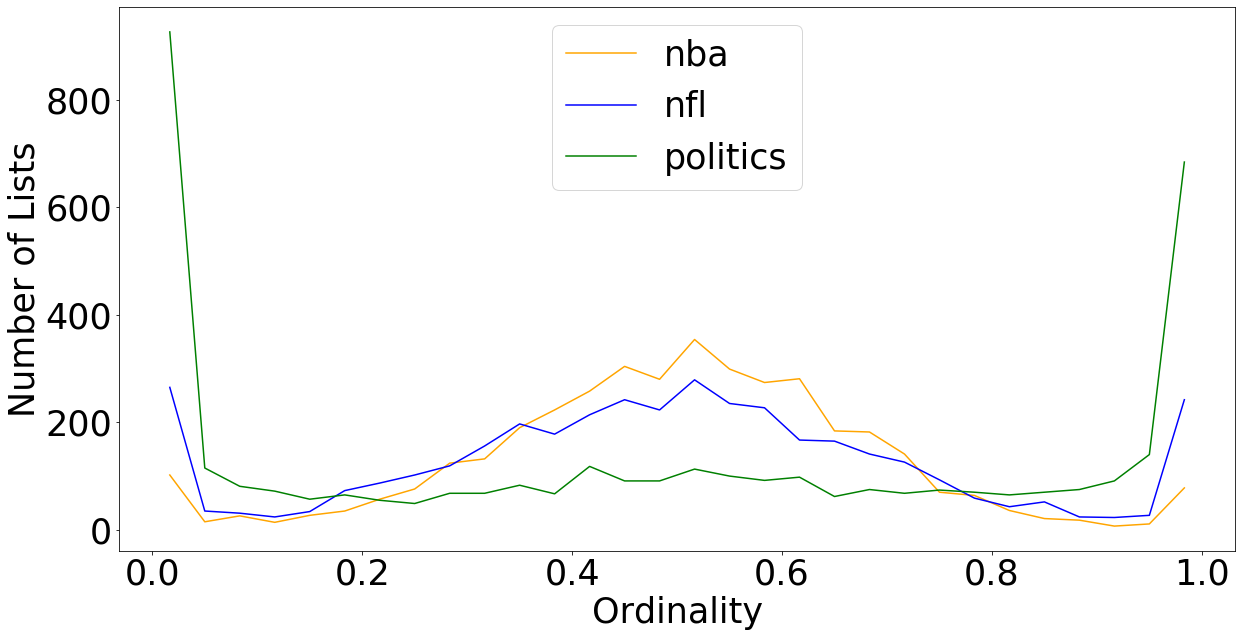}
    \caption{Histogram of asymmetry for lists of names in r/nfl, r/nba and r/politics.}
     
    \label{fig:asymmetry hist names}
\end{figure}

\subsection{NBA Names}
First, we examined names in r/nba. One advantage of using NBA players is that we have detailed statistics for ever player in every year. We tested a number of these statistics, and while all of them predicted statistically significant numbers ($p <$ 1e-6) 
of binomials,
they were still not very  predictive in a practical sense (Table~\ref{tab: sports results}). The best predictor was actually how often the player's team was mentioned. Interestingly, the unigram frequency (number of times the player's name was mentioned overall) was not a good predictor.
It is relevant to these observations that some team subreddits (and thus, presumably, fanbases) are significantly larger than others.

\begin{table}[]
\caption{Prediction results for a number of statistics about players in the NBA and NFL.
A `-' denotes that the statistic is unavailable for the sport.}
\label{tab: sports results}
\begin{tabular}{lll}
\toprule
              & r/nba & r/nfl \\
              \midrule
Alphabetical  & 0.49 & 0.48       \\
Num Chars     & 0.40 & 0.40      \\
Unigram Freq  & 0.50 & 0.49      \\
Years Playing & 0.56 & 0.65      \\
Team Mentions & 0.70 & 0.83      \\
Games Played  & 0.54 & 0.62   \\
PER           & 0.51 & -      \\
VORP          & 0.51 & -     \\
Games Started & - & 0.60 \\
Linear Model  & 0.56  & 0.51 \\     
\bottomrule
\end{tabular}
\end{table}

\subsection{Subreddit and team names}
Additionally, we also investigated lists of names of sports teams and subreddits as proper nouns. In this case we exploit an interesting structure of the r/nba subreddit which is not evident at scale in other subreddits we examined. In addition to r/nba, there exists a number of subreddits that are affiliated with a particular NBA team, with the purpose of allowing discussion between fans of that team. This implies that most users in a team subreddit are fans of that team.
We are then able to look for lists of NBA teams by name, city, and abbreviation. We found 2520 instances of the subreddit team coming first, and 1894 instances of the subreddit team coming second. While this is not a particularly strong predictor, correctly predicting 57\% of lists, it is one of the strongest we found, and a clear illustration of the Proximity Principle.

We can do a similar calculation with subreddit names, by looking
between subreddits. While the team subreddits are not large enough for
this calculation, many of the other subreddits are.
We find that lists of subreddits in r/nba that include `r/nba' often start with `r/nba',
and a similar result holds for r/nfl (Table~\ref{tab: sport v sport}).


\begin{table}[]
\caption{If two sports subreddits are listed in a sports subreddit, the subreddit of origin (r/nba in top row, r/nfl in bottom row) usually comes first, in terms of the weighted token evaluation (number of occurrences in parentheses). 
A `-' means that there are fewer than 30 such lists.}
\label{tab: sport v sport}
\begin{tabular}{llllll}
\toprule
      & r/nba     & r/nfl     & r/hockey & r/soccer & r/baseball \\ 
      \midrule
r/nba & -         & 0.73 (204) & 0.73 (34) & 0.44 (43) & -          \\ 
r/nfl & 0.61 (204) & -         & 0.66 (35) & 0.60 (75)  & 0.76 (30)   \\ 
\bottomrule
\end{tabular}
\end{table}

\omt{
\begin{table}[]
\caption{If two politics subreddits are listed in a politics subreddit, the subreddit of origin sometimes comes first, but r/politics is almost always first}
\label{tab: politics v politics}
\begin{tabular}{lllll}
\toprule
               & r/Conservative & r/Libertarian & r/politics & r/The\_Donald \\ 
\midrule
r/Conservative & -              & 0.51(35)       & 0.44(104)   & -             \\ 
r/Libertarian  & 0.49(78)        & -             & 0.48(82)    & -             \\ 
r/politics     & 0.66(89)       & 0.61(36)      & -          & 0.78(187)     \\ 
r/The\_Donald  & -              & -             & 0.47(34)    & -             \\ 
\bottomrule
\end{tabular}
\end{table}
}

While NBA team subreddits show a fairly strong preference to name
themselves first, this preference is slightly less strong among sport
subreddits, and even less strong among politics subreddits. One
potential factor here is that r/politics is a more general subreddit, while
the rest are more specific --- perhaps akin to r/nba and the team
subreddits. 

\subsection{Political Names}
In our case, political names are drawn from every congressperson (and their nicknames) in both houses of the US Congress through the 2018 election. It is worth noting that one of these people is Philadelph Van Trump. It is presumed that most references to `trump' refer to  Donald Trump. There may be additional instances of mistaken identities. We restrict the names to only congresspeople that served before 1801 or after 1999, also including `trump'.

\begin{table}[tb]
\caption{Political name ordering by party across political subreddits.
Note that r/politics is left-leaning.}
    \begin{tabular}{l*{5}c}
    \toprule
         & (D,D) & (D,R) & (R,D) & (R,R) \\
         \midrule
         r/politics & 3167 & 2263 & 1980 & 4442 \\
         r/Conservative & 313 & 260 & 225 & 610 \\
         r/Libertarian & 342 & 311 & 235 & 612 \\
         \bottomrule
    \end{tabular}
    \label{tab:party order}
\end{table}

One might guess that political subreddits refer to politicians of their preferred party first. 
However, this was not the case, as Republicans are mentioned first only about 43\%--46\% of the time in all subreddits (Table~\ref{tab:party order}).
On the other hand, the Proximity Principle does seem to come into play when discussing ideology. For instance, r/politics --- a left-leaning subreddit --- is more likely to say `democrats and republicans' while the other political subreddits  in our study --- which are right-leaning --- are more likely to say `republicans and democrats'. 


Another relevant measure for lists of proper nouns is the ratio of the number of list instances containing a name to the unigram frequency of that name.
We restrict our investigation to names that are not also English words, and only names that have a unigram frequency of at least 30.
The average ratio is 0.0535, but there is significant variation across names. It is conceivable that this list ratio is revealing about how often people are talked about alone instead of in company.

\omt{
\begin{table}[]
    \centering
    \caption{Most unusual lists and names by subreddit compared to unweighted average percentage among politics subreddits.}
    \begin{tabular}{l| p{6cm}}
        \toprule
        \multicolumn{2}{c}{Unusual Lists} \\
        \midrule
         politics & ['engl', 'don'], ['obama', 'nancy pelosi'], ['lee atwater', 'karl'] \\
         Conservative & ['kathy griffin', 'holding'], ['rubio', 'carson'], ['english', 'french'] \\
         Libertarian & ['gary', 'bill'], ['gary', 'jill'], ['laws', 'law'] \\
         The\_Donald & ['patton', 'macarthur'], ['trump', 'ron'], ['trump', 'new'] \\
         \midrule
         \multicolumn{2}{c}{Unusual Names} \\
        \midrule
        politics & 'more', 'trump', 'obama \\
        Conservative & 'more', 'trump', 'cruz' \\
        Libertarian & 'paul', 'r', 'more' \\
        The\_Donald & 'more', 'trump', 'hillary' \\
        \bottomrule
    \end{tabular}
    \label{tab:unique lists}
\end{table}
}

\omt{
\begin{figure}[h!]
    \includegraphics[width=.45\textwidth]{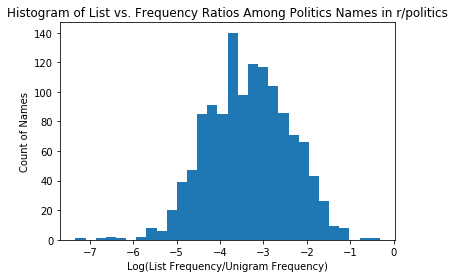}
  \caption{Histogram of log(list frequency/unigram frequency) for political names in r/politics. This shows that the ratio varies quite a bit from one name to another.}
  \label{fig: list ratio plot}
\end{figure}
}

\section{Formal Text}
While Reddit provides a very large corpus of informal text, McGuire and McGuire make a distinct separation between informal and formal text \cite{kinpairs}. As such, we briefly analyze highly stylized wine reviews and news articles from a diverse set of publications. Both data sets follow the same basic principles outlined above.

\begin{table}[]
\caption{Number of total lists (log scale) and percent of lists that are frozen. There is no correlation between size and frozenness, but note that news is far more frozen than any other data source.}
\label{tab:frozenness}
\begin{tabular}{@{}lll@{}}
\toprule
           & Size & Frozenness \\ \midrule
news       & 13.1 & 0.41       \\
r/food     & 12.6 & 0.24       \\
f/nba      & 14.6 & 0.16       \\
r/politics & 16.4 & 0.28       \\
wine       & 11.1 & 0.25       \\ \bottomrule
\end{tabular}
\end{table}

\subsection{Wine}
Wine reviews are a highly stylized form of text. In this case reviews are often just a few sentences, and they use a specialized vocabulary meant for wine tasting. While one might hypothesize that such stylized text exhibits more frozen binomials, this is not the case (Tab~\ref{tab:frozenness}). There is some evidence of an additional freezing effect in binomials such as ('aromas', 'flavors') and ('scents', 'flavors') which both are frozen in the wine reviews, but are not frozen on Reddit. However, this does not seem to have a more general effect. Additionally, there are a number of binomials which appear frozen on Reddit, but have low asymmetry in the wine reviews, such as ['lemon', 'lime']. 

\subsection{News}
We focused our analysis on NYT, Buzzfeed, Reuters, CNN, the Washington Post, NPR, Breitbart, and the Atlantic. Much like in political subreddits, one might expect to see a split between various publications based upon ideology. However, this is not obviously the case. While there are certainly examples of binomials that seem to differ significantly for one publication or for a group of publications (Buzzfeed, in particular, frequently goes against the grain), there does not seem to be a sharp divide. Individual examples are difficult to draw conclusions from, but can suggest trends. (`China', `Russia') is a particularly controversial binomial. While the publications vary quite a bit, only Breitbart has an ordinality of above 0.5. In fact, country pairs are among the most controversial binomials within the publications (e.g. (`iraq', `syria'), (`afghanisatan', `iraq')), while most other highly controversial binomials reflect other political structures, such as (`house', `senate'), (`migrants', 'refugees'), and (`left', `right'). That so many controversial binomials reflect politics could point to subtle political or ideological differences between the publications. Additionally, the close similarity between Breitbart and more mainstream publications could be due to a similar effect we saw with r/The\_Donald - mainly large amounts of quoted text.

\section{Global Structure}
\begin{figure}[t]
    \includegraphics[width=.45\textwidth]{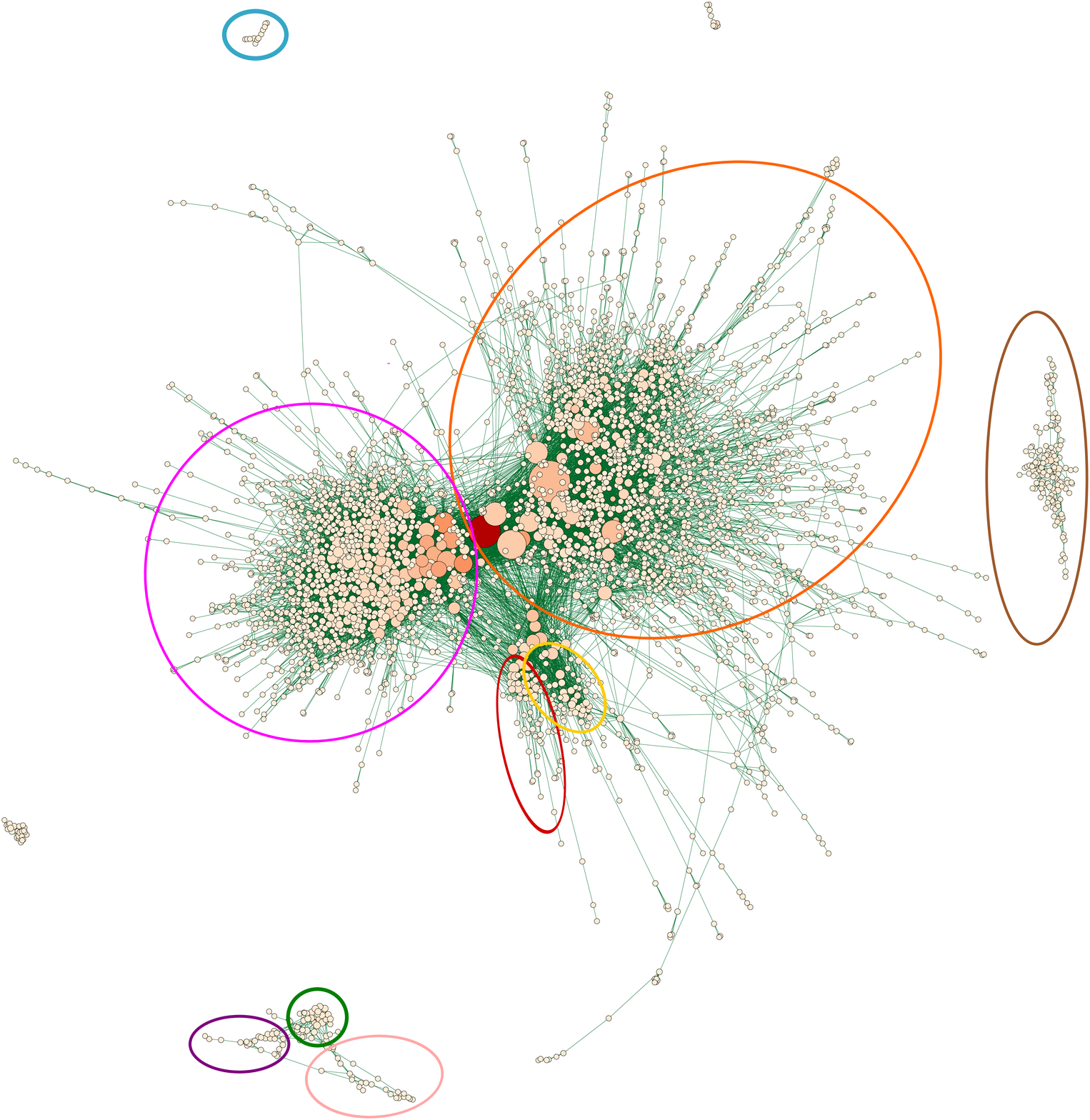}
    \caption{The r/nba binomial graph, where nodes are words and directed edges indicate binomial orientation.}
    \label{fig: NBA graph}
\end{figure}

\begin{figure}[t]
    \includegraphics[width=.45\textwidth]{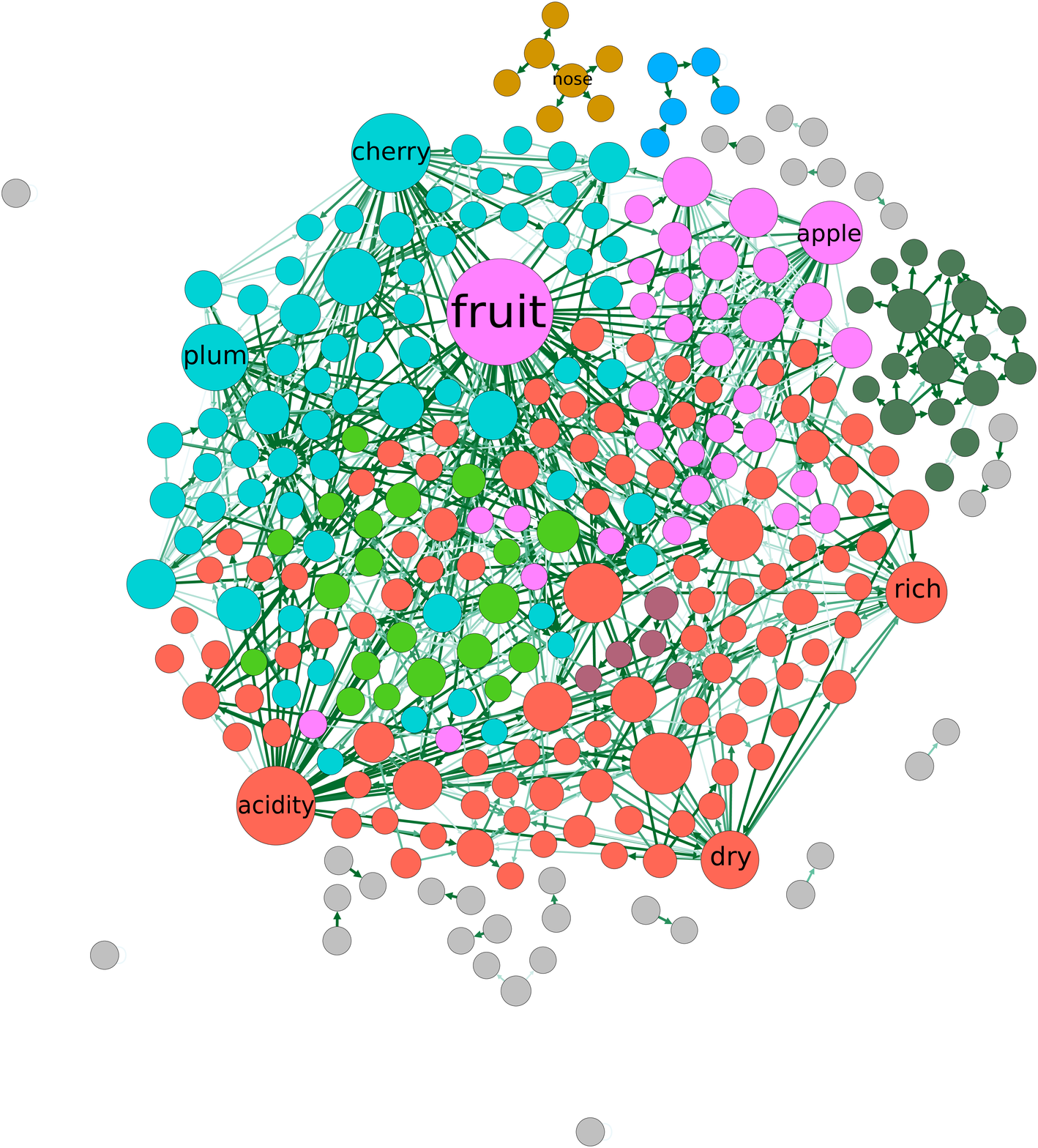}
    \caption{The wines binomial graph, where nodes are words and directed edges indicate binomial orientation.}
    \label{fig: wine graph}
\end{figure}

We can discover new structure in binomial orderings by taking a more global view. We do this by building directed graphs based on ordinality. 
In these graphs, nodes are words and an arrow from A to B indicates that there are at least 30 lists containing A and B and that those lists have order [A,B] at least 50\% of the time.
For our visualizations, the size of the node indicates how many distinct lists the word appears in,and color indicates how many list instances contain the word in total.

If we examine the global structure for r/nba, we can pinpoint a number of patterns (Fig.~\ref{fig: NBA graph}). First, most nodes within the purple circle correspond to names, while most nodes outside of it are not names. The cluster of circles in the lower left are a combination of numbers and years, where dark green corresponds to numbers, purple corresponds to years, and pink corresponds years represented as two-digit numbers (e.g., `96'). On the right, the brown circle contains adjectives, while above the blue circle contains heights (e.g., 6'5"), and in the two circles in the lower middle, the left contains cities while the right contains team names. The darkest red node in the center of the graph corresponds to `lebron'. 

Constructing a similar graph for our wines dataset, we can see clusters of words. In Fig~\ref{fig: wine graph}, the colors represent clusters as formed through modularity. These clusters are quite distinct. Green nodes mostly refer to the structure or body of a wine, red are adjectives describing taste, teal and purple are fruits, dark green is wine varietals, gold is senses, and light blue is time (e.g. `year', `decade', etc.)

We can also consider the graph as we change the threshold of asymmetry for which an edge is included.
If the asymmetry is large enough, the graph is acyclic, and we can consider how small the ordinality threshold
must be in order to introduce a cycle. 
These cycles reveal the non-global ordering of binomials.
The graph for r/nba begins to show cycles with a threshold asymmetry of 0.97. 
Three cycles exist at this threshold: [`ball', `catch', `shooter'], [`court', `pass', `set', `athleticism'], and [`court', `plays', `set', `athleticism'].

Restricting the nodes to be names is also revealing.
Acyclic graphs in this context suggest a global partial hierarchy of individuals.
For r/nba, the graph is no longer acyclic at an asymmetry threshold of 0.76,
with the cycle [`blake', `jordan', `bryant', `kobe'].
Similarly, the graph for r/nfl (only including names) is acyclic until the threshold reaches 0.73 with cycles 
[`tannehill', `miller', `jj watt', `aaron rodgers', `brady'], and [`hoyer', `savage', `watson', `hopkins', `miller', `jj watt', `aaron rodgers', `brady'].

\omt{
\subsection{Interesting Examples}
\subsubsection{Political Names}
One interesting application of these graphs are examining names of politicians within different politics subreddits. Examining these graphs can reveal differences in these communities in who they talk about and who they put first. If we accept some of the Proximity Principle and Me First Principle~\cite{cooperandross}, one way to understand these differences in order is who the community relates with more. For instance, Obama and Trump are central to all three graphs, but Cruz only appears in r/Conservative as central. Merely examining the lists present can also reveal who is being compared and discussed together. [obama, biden] and [bush, cheney] are both quite common, unsurprisingly.
}

\begin{figure*}
    \centering
    \begin{subfigure}{.3\textwidth}
        \includegraphics[width=.9\textwidth]{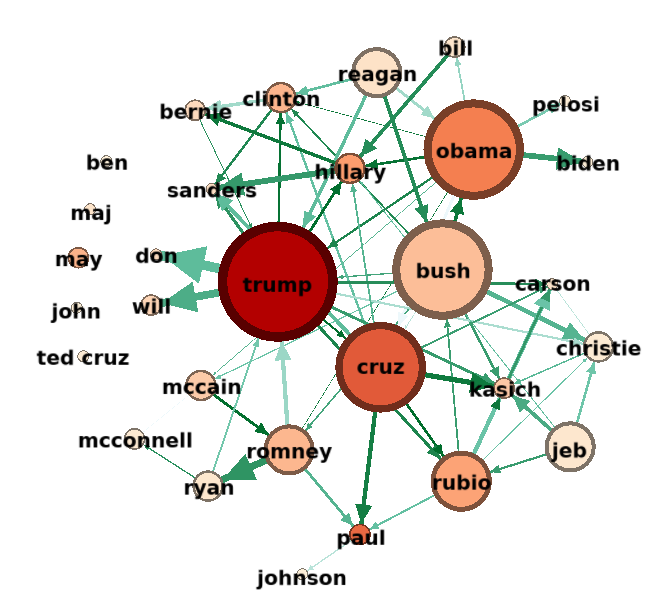} 
        \caption{r/Conservative}
    \end{subfigure}\begin{subfigure}{.3\textwidth}
        \includegraphics[width=.9\textwidth]{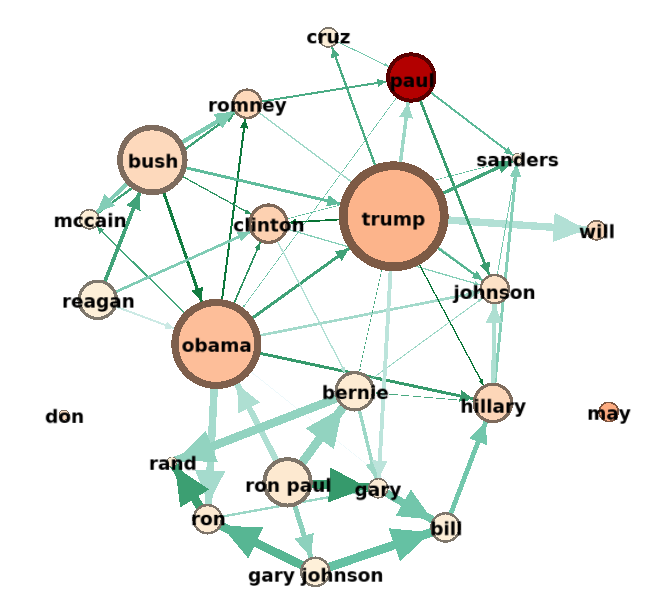} 
        \caption{r/Libertarian}
    \end{subfigure}\begin{subfigure}{.3\textwidth}
        \includegraphics[width=.9\textwidth]{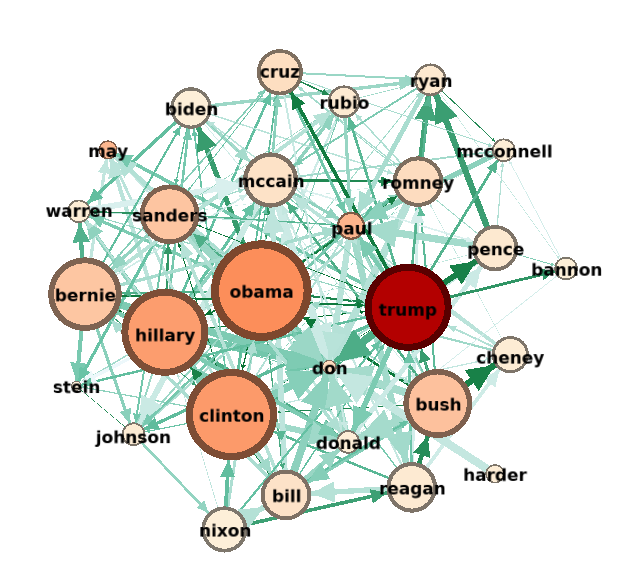}
        \caption{r/politics}
    \end{subfigure}
    \caption{Graphs of some of the 30 most common names in r/Conservative, r/Libertarian, and r/politics. 
    Nodes are names, and an edge from A to B represents a list where the dominant order is [A,B]. Node size is the number of lists the word comes first in, node color is the total number of lists the node shows up in, edge color is the asymmetry of the list.}
     
    \label{fig: politics names}
\end{figure*}


Figure~\ref{fig: politics names} shows these graphs for the three political subreddits, where the nodes
are the 30 most common politician names.
The graph visualizations immediately show that these communities view politicians differently.
We can also consider cycles in these graphs and find that the
graph is completely acyclic when the asymmetry threshold is at least 0.9.
Again, this suggests that, at least among frozen binomials, there is in fact a global partial order of names that might signal hierarchy.
(Including non-names, though, causes the r/politics graph to never be acyclic for any asymmetry threshold, 
since the cycle [`furious', `benghazi', `fast'] consists of completely frozen binomials.)
We find similar results for r/Conservative and r/Libertarian, which are acyclic with thresholds of 0.58 and 0.66, respectively. 
Some of these cycles at high asymmetry might be due to English words that are also names (e.g. `law'),
but one particularly notable cycle from r/Conservative is [`rubio', `bush', `obama', `trump', `cruz'].

\section{Multinomials}
Binomials are the most studied type of list, but trinomials --- lists of three --- are also common enough in our dataset to analyze.
Studying trinomials adds new aspects to the set of questions: for example, while binomials have only two possible orderings, trinomials have six possible orderings.
However, very few trinomials show up in all six orderings. 
In fact, many trinomials show up in exactly one ordering: about 36\% of trinomials being completely frozen amongst trinomials appearing at least 30 times in the data.
To get a baseline comparison, we found an equal number of the most common binomials, and then subsampled instances of those binomials to equate the number of instances with the trinomials. 
In this case, only 21\% of binomials are frozen.
For trinomials that show up in at least two orderings, it is most common for the last word to keep the same position (e.g., [a, b, c] and [b, a, c]). 
For example, in our data, [`fraud', `waste', `abuse'] appears 34 times, and [`waste', `fraud', `abuse'] appears 20 times. 
This may partially be explained by many lists that contain words such as `other', `whatever', or `more'; for instance,
[`smarter', `better', `more'] and [`better', `smarter', `more'] are the only two orderings we observe for this set of three words.

Additionally, each trinomial [a, b, c] contains three binomials within it: [a, b], [b, c], and [a, c].
It is natural to compare orderings of \{a, b\} in general with orderings of occurrences of \{a, b\} that lie inside trinomials.  We use this comparison to define the {\em compatibility} of \{a, b\}, as follows.

\begin{definition}{\textbf{Compatibility}}
Let \{a, b\} be a binomial with dominant ordering [a, b]; that is, [a, b] is at least as frequent as [b, a]. We define the {\em compatibility} of \{a, b\} to be the fraction of instances of \{a, b\} occurring inside trinomials that have the order [a,b].
\end{definition}

\begin{figure}[h!]
    \includegraphics[width=.35\textwidth]{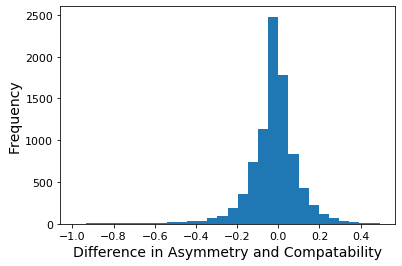}
    \caption{Histogram of difference in asymmetry and compatibility for binomials within trinomials on r/politics.}
     
    \label{fig:trinomial diff}
\end{figure}

There are only a few cases where binomials have compatibility less than 0.5, and
for most binomials, the  
asymmetry is remarkably consistent between binomials and trinomials (Fig.~\ref{fig:trinomial diff}).
In general, asymmetry is larger than compatibility --- this occurs for 4569 binomials,
compared to 3575 where compatibility was greater and 690 where the two values are the same.
An extreme example is the binomial \{`fairness', `accuracy'\}, which has asymmetry 0.77 and compatibility 0.22.
It would be natural to consider these questions for
tetranomials and longer lists, but these are
rarer in our data and correspondingly harder to draw conclusions from.

\omt{
\begin{figure}[h!]
    \includegraphics[width=.45\textwidth]{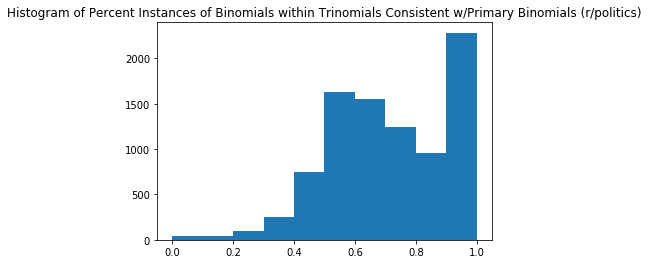}
    \caption{Histogram of the compatibility of binomials found within trinomials in r/politics}
    \label{fig: trinomial compatibility}
\end{figure}

\begin{figure}[h!]
    \includegraphics[width=.45\textwidth]{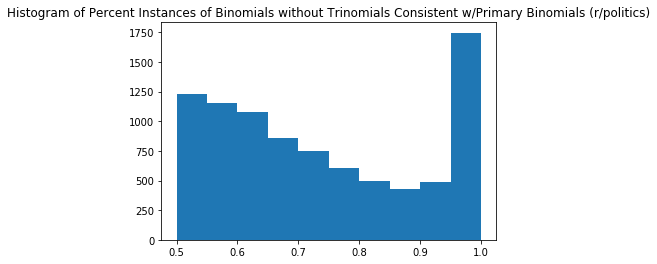}
    \caption{Histogram of the asymmetry of the binomials found in trinomials in r/politics}
    \label{fig:trinomial asymmetry}
\end{figure}
}

\section{Discussion}
Analyzing binomial orderings on a large scale has led to surprising results. 
Although most binomials are not frozen in the traditional sense, 
there is little movement in their ordinality across time or communities. 
A list that appears in the order [A, B] 60\% of the time in one subreddit in one year is likely to show up as [A, B] very close to 60\% of the time in all subreddits in all years. 
This suggests that binomial order should be predictable, but there is evidence that this is difficult:
the most common theories on frozen binomial ordering were largely ineffective at predicting binomial ordering in general.

Given the challenge in predicting orderings, we searched for methods or principles that could yield better performance, and identified two promising approaches.
First, models built on standard word embeddings produce predictions of binomial orders that are much more effective than simpler existing theories.
Second, we established the Proximity Principle: the proper noun with which a speaker identifies more will tend to come first. This is evidenced when commenters refer to their sports team first, or politicians refer to their party first. Further analysis of the global structure of binomials reveals interesting patterns and a surprising acyclic nature in names.
Analysis of longer lists in the form of multinomials suggests that the rules governing their orders may be different. 

We have also found promising results in some special cases.
We expect that more domain-specific studies will offer rich structure.

It is a challenge to adapt the long history of work on the question of frozen binomials to the large, messy environment of online text and social media. However, such data sources offer a unique opportunity to re-explore and redefine these questions. It seems that binomial orderings offer new insights into language, culture, and human cognition. Understanding what changes
in these highly stable conventions mean --- and whether or not they can be predicted --- is an interesting avenue for future research.

\section{Acknowledgements}
The  authors  thank  members  of  the Cornell AI, Policy, and Practice Group, and (alphabetically by first name) Cristian Danescu-Niculescu-Mizil, Ian Lomeli, Justine Zhang, and Kate Donahue for aid in accessing data and their thoughtful insight.
This research was supported by
NSF Award DMS-1830274,
ARO Award W911NF19-1-0057,
a Simons Investigator Award,
a Vannevar Bush Faculty Fellowship,
and ARO MURI.

\omt{
\begin{table}[]
\caption{The food hierarchy predicts what sorts of foods come before others in binomials (fish, meat, drink, fruit, vegetables, baked goods, dairy products, spices). We tested this using data from the subreddit r/food and extracting lists of foods by giving ~10-20 words that were examples of each category, and then testing to see if each list followed the hierarchy. }
\label{tab:food}
\begin{tabular}{|l|l|}
\hline
W. Token & 0.59 \\ \hline
U. Token & 0.57 \\ \hline
W. Type  & 0.68 \\ \hline
U. Type  & 0.64 \\ \hline
\end{tabular}
\end{table}
}

\bibliographystyle{aaai}
\bibliography{refs}
\end{document}